**Irene S. Gabashvili, Christopher K. Allsup**
Aurametrix, USA
https://aurametrix.com

# Resident Turnover and Community Satisfaction in Active Lifestyle Communities

## Abstract

Despite the growing significance of active adult lifestyle communities, systematic econometric studies on resident tenure and satisfaction remain scarce. Using longitudinal surveys (2018, 2021, 2023, 2024) and property transaction records, we analyze demographic shifts, homeownership tenure, resident satisfaction, and departure patterns in Tellico Village through survival analysis, hazard rate modeling, and machine learning techniques.

Large-scale surveys report high satisfaction levels (93%), well above comparable communities (~73%), with an average homeownership duration of 14-15 years. However, property records reveal a lower median tenure of 11 years, down from 13 years pre-COVID (Log-Rank Test: $p < 0.001$, $-\log_2(p) = 72$), reflecting a broader nationwide trend in homeownership duration. Post-COVID, tenure varies significantly by neighborhood, ranging from 8 to 15 years. Kaplan-Meier survival analysis and Weibull hazard modeling identify departure risk peaks at 3, 5, 7, 11, 16, 22, and 26 years, reflecting personal life transitions, housing market cycles, and community adaptation processes. The two-component Weibull mixture model captures distinct tenure subgroups, with shape ($k$) and scale ($\lambda$) parameters optimized via non-linear fitting, outperforming alternative hazard models (Gaussian, Lorentzian, Exponential, Voigt).

Resident satisfaction follows a U-shaped trajectory, reaching its lowest point between years 3–12, aligning with departure risk peaks at years 3, 5, 7, and 11. Regression analysis, after standardizing predictors for comparability, identifies financial attitudes, recreational engagement, and openness to growth as key factors associated with higher satisfaction.

While willingness to pay higher POA fees exhibits a strong linear relationship with overall satisfaction ($R^2 = 94\%$, explaining 94% of the variance) at the aggregate level, accounting for 44% of the variance at the neighborhood level and only 5% at the household level, underscoring the role of additional factors. Newer and younger residents demonstrate greater price sensitivity, influencing governance preferences and amenity valuation. Machine learning models—including Support Vector Machines, Random Forests, and XGBoost—were employed to predict satisfaction and tenure, but latent variables and class imbalances limited their predictive accuracy. Hedonic price modeling, leveraging property appraisals and appreciation trends, provided insights into tenure behaviors at the neighborhood level but was constrained by the absence of individual household financial data.

To address survey bias and potential response distortions, Bayesian post-stratification weighting and comparative statistical analyses—including Kruskal-Wallis, ANOVA, Mann-Whitney U tests, and hierarchical modeling—were applied to assess generational and neighborhood-level variations. Additionally, the introduction of a COVID Impact Index (CII), weighted by Bayesian posterior probability, enabled the assessment of pandemic-driven tenure shifts.

Our findings contribute to the econometrics of housing tenure, satisfaction, and community stability, demonstrating the complex interplay of demographic shifts, housing market fluctuations, and governance choices. Future research should incorporate instrumental variable approaches, longitudinal tracking, and improved feature engineering in predictive modeling to refine tenure and satisfaction forecasts in master-planned communities.

## Introduction

### Background

Active adult communities are reshaping American residential patterns, yet their internal dynamics remain poorly understood. Designed primarily but not exclusively for residents aged 55 and older, these neighborhoods offer strong social connections and a wide range of amenities that promote an engaged way of life. However, shifting demographics, economic pressures, and broader societal changes—especially in the post-COVID-19 era—are altering how residents perceive long-term community engagement.

Resident turnover and satisfaction are central to the long-term success of these communities. While some homeowners settle permanently, others move out within a few years due to financial, social, health-related or lifestyle changes. Understanding these patterns is crucial, as high turnover can affect community stability, governance, and property values. Additionally, shifting resident expectations—whether regarding amenities, affordability, or governance—can shape the future development of these communities.

Despite the growing prevalence of active adult communities, systematic research on their long-term dynamics is scarce. While national housing data from sources like Redfin [1] and the U.S. Census [2] provide insights into general homeownership trends, they rarely differentiate between traditional housing and active adult communities. This lack of segmentation obscures critical patterns in turnover, satisfaction, and community evolution.

Part of this research gap stems from the fluid definition of "active adult" communities [3]. Unlike senior living models such as Independent Living or Assisted Living, active adult communities operate as a market-driven concept rather than a regulated housing category [4]. They vary widely in amenities, infrastructure, and governance structures, making comparative studies challenging.

Existing research highlights key issues but remains limited in scope. For example, surveys indicate high resident satisfaction—73% in a study of 873 active adult rental community residents across 37 states [5]. However, long-term concerns such as infrastructure, accessibility, and generational shifts remain underexplored. Studies suggest that early-stage communities benefit from strong identity formation [6], but satisfaction over time depends on governance adaptability and evolving resident priorities.

Recent work on modeling construction growth in master-planned communities [7] underscores the importance of dynamic ownership patterns. Additionally, strategic planning tools designed to quantify resident preferences [8] highlight the role of subjective priorities in long-term community planning. These insights point to the need for a more integrated approach—one that considers evolving demographics, shifting resident priorities, and economic forces shaping active adult communities.

### Rationale

This study fills a critical gap by analyzing homeownership tenure dynamics, resident satisfaction trajectories, departure trends, and financial correlates within an active adult community. Using large-scale surveys and official property records, we evaluate and refine predictive models to enhance strategic planning for future developments.

Building on prior work [7], this work applies similar property ownership and transaction data sources but shifts the focus from undeveloped lots to completed homes, offering new insights into built-environment dynamics within active adult communities.

### Methods

### Data acquisition and transformation

*Data Sources*

The data analyzed in this study were drawn from multiple sources to provide a comprehensive overview of resident satisfaction, demographic trends, and community engagement across different time periods:

• 2018 Survey of Tellico Village Residents (n = 1674): Captured pre-pandemic baseline satisfaction and demographics.
• 2021 Survey of Tellico Village Residents (n = 2335): Reflected mid-pandemic changes in attitudes and move-out intentions.
• 2023 Survey of Restaurants (n = 2725): Conducted at the end of 2023 to explore residents' dining habits, preferences, and frequency of visits to POA-run restaurants, while also collecting demographic data (neighborhood and age).
• 2024 Surveys:
    - Meeting Spaces (n = 573): Focused on amenity usage and engagement across different neighborhoods.
    - Parking (n = 1178): Examined parking challenges and usage patterns.
• Home Sales Records from Property Appraiser's Database (1984–2025): A dataset of 43,887 transactions capturing actual move-out events, durations of property and homeownership, appraisals, square footage, and features such as decks and boat slips. Data were sourced from the Tennessee Comptroller of the Treasury's [Tennessee Property Assessment Data](#) (TPAD), covering two jurisdictions and the eight neighborhoods that comprise Tellico Village. [7]

For broader population-based estimates, we utilized:

• U.S. Census Bureau's American Community Survey (ACS): Provided data on disability rates. [9]
• National Center for Health Statistics (NCHS) of the Centers for Disease Control and Prevention (CDC): Supplied mortality statistics [10].
• Tellico Village health data: Collected as previously described [11].

### *Key Variables*

- Satisfaction Scores: Overall community satisfaction, aesthetics satisfaction, amenity satisfaction (e.g., golf, restaurants, recreational facilities), satisfaction with community growth and development.
- Demographics: Household Size and age distribution of all occupants
- Price Sensitivity: Measured through direct questions about willingness-to-pay and acceptance of fee increases.
- Tenure and Housing Dynamics: Years of residence in the community at the time of the survey, housing type and ownership status, future downsizing plans
- Community Engagement and Participation: Frequency of amenity usage and participation in social activities, ranking and perceived importance of community amenities, experiences with events (including frequency of capacity issues affecting attendance at events, dinners, or recreation)

- Transportation: Experiences with transportation and parking, carpooling behaviors and preferences
- Employment and Civic Engagement: employment status and the need to commute, levels of civic participation
- Neighborhood and Location attributes: Proximity to amenities (using property records and geographic data), frequency and nature of neighborhood-specific events and activities
- Health and Nutrition: Evaluated restaurant survey responses for suggested menu options and their healthiness, identified neighborhoods with statistically significant healthier and unhealthier scores based on dietary preferences and choices.

### *Data Cleaning and Transformation*

Several data transformation techniques were applied. Ordinal encoding was used for responses on 5-point and 3-point Likert scales, while one-hot encoding was applied to categorical variables such as "Yes" and "No," converting them into a binary (0 or 1) matrix representation. Frequency encoding was employed for questions regarding amenity usage. To address high dimensionality, Principal Component Analysis (PCA) was explored on one-hot encoded data.

Missing values were infrequent, and unless otherwise stated, rows with missing values were removed - for example, responses on household tenure, which were optional for renters and non-property owners. However, in some analyses, missing entries were imputed while preserving the data structure. This included using category-specific averages (e.g., renters) or neighborhood-level imputation, applying mode imputation for categorical variables and median imputation for numerical variables.

### *Feature Engineering*

We created composite indicators and labels based on survey responses. For example, we combined golfers and wellness center memberships into an "active recreation interest" score and classified respondents as "Urban Convenience Seekers" or "Nature Enthusiasts" based on amenity preferences and environmental values. Financial preferences were categorized into groups such as "Strict Cost Cutters," "Moderate Investors," and "Prudent Planners," reflecting attitudes toward community funding and development.

Another approach involved segmenting residents based on funding preferences for future amenities and maintenance. Labels such as "Establishment Supporters," "Amenity Maximizers," and "Revenue Diversifiers" were created based on responses to different fee and funding strategies. Additionally, we combined two key dimensions—support for fee minimization and support for necessary increases—into a single latent attitude measure, "willingness to pay higher POA dues" score, allowing us to distinguish between moderate and extreme positions on funding policies.

Interaction terms were generated to capture the relationship between household demographics and facility usage, such as the number of residents in a household combined with interest in additional amenities or price sensitivity linked to a neighborhood's appraisal-based cost index. The structured data enabled multi-stage analysis, enhancing our ability to explore community trends and behaviors.

**Statistical and Analytical Techniques**

Our analysis of residential tenure patterns employed the Kaplan-Meier survival method to estimate how long residents remained in the overall community and across its eight neighborhoods, with a focus on pre- and post-pandemic shifts. Key survival metrics included median tenure (MT), daily and yearly survival rates, and the cumulative hazard function, which estimated move-out risk over time. To assess statistical significance, we used bootstrap-computed confidence intervals and log-rank tests.

To quantify the impact of COVID-19, we developed a COVID Impact Index (CII), derived from differences in median tenure before and after the pandemic, weighted by posterior probability. The Bayesian framework allowed for the incorporation of effect size and uncertainty, ensuring a more nuanced evaluation than reliance on p-values alone.

$$CII = \left(\frac{PreCOVID\ MT - PostCOVID\ MT}{PreCOVID\ MT}\right) * PosteriorWeight$$

Given that p-values are highly sensitive to sample size—causing smaller neighborhoods to exhibit weaker statistical significance even for meaningful differences—we prioritized standardized effect sizes and Bayesian metrics to provide a more balanced interpretation.

To establish a baseline for tenure patterns, we generated histograms of genuine sales transactions, filtering out legal transfers that did not reflect actual relocations. This revealed key patterns in tenure length distributions.

Our exploration of hazard rate modeling identified the Weibull hazard function as the best fit, capturing distinct subgroups of short- and long-term residents through two-component Weibull mixtures. Model parameters—shape (k) and scale (λ)—were determined via non-linear optimization, outperforming alternative models such as Gaussian, Lorentzian, Exponential, and Voigt distributions.
Each Weibull function follows the form:

$$f(t) = \frac{k}{\lambda}\left(\frac{t}{\lambda}\right)^{k-1} e^{-(\frac{t}{\lambda})^k}$$

where t is tenure duration, k is shape parameter, indicating whether the likelihood of leaving increases or decreases over time and $\lambda$ is scale parameter, which adjusts the spread of the distribution

To examine temporal trends, we segmented the data into distinct periods, comparing tenure patterns before and after major economic and social events. By integrating Kaplan-Meier survival analysis with hazard rate calculations, we identified key departure risk periods and determined when the probability of moving out exceeded the average.

To address volatility in daily departure rates, we calculated annualized hazard rates, stabilizing short-term fluctuations and revealing long-term movement patterns. This methodology provided a clearer picture of sustained tenure trends.

### Comparative and Predictive Analyses

To gain deeper insights into resident satisfaction and turnover, we applied a combination of statistical tests and machine learning methods. When the Shapiro-Wilk test indicated non-normal distributions, we used the Kruskal-Wallis H test and Mann-Whitney U test for non-parametric comparisons. For normally distributed data, we conducted t-tests and ANOVA, followed by Tukey's HSD for post-hoc analysis.

For predictive modeling, we employed Random Forest and Linear Regression to assess overall performance, using the coefficient of determination ($R^2$) as a measure of predictive strength. LASSO regression aided in feature selection, while Support Vector Machines (SVM) with class weighting addressed class imbalance. Hierarchical modeling captured neighborhood-level variations in tenure and satisfaction, and decision trees optimized tenure binning based on statistically significant split points from t-tests. Although XGBoost was hyperparameter-tuned, its performance was constrained by the dataset structure. Additionally, hedonic regression was used to quantify the impact of property and market characteristics on satisfaction, recommitment (measured by the survey question: "Would you still choose this community if you had to decide again?"), and residential tenure.

Transformer-based models, including ChatGPT o1 and o3-mini-high, were leveraged to classify and analyze verbatim comments. To mitigate survey response biases, post-stratification weighting adjusted for neighborhood over- and underrepresentation. While a Heckman selection model would typically correct for response bias, data limitations required careful interpretation of potential distortions.

### Results and Discussion

#### Resident Demographics

Tellico Village's resident profile is defined by a predominantly mature population, with evolving household compositions compared to earlier surveys. The age distribution (N = 4,567) shows that over two-thirds of residents fall within the 65–79 age range, collectively comprising approximately 65.7% of the community. In contrast, younger age groups remained underrepresented, with only 261 residents under 55 (5%). This pronounced age skew reinforces the community's identity as an active adult lifestyle destination.

Despite the community's aging profile, recent data suggest increasing household diversity. However, much of the available statistical data still reflects the 2021 survey. In that year:
- Baby Boomers: The largest segment, accounting for 1,469 households, with most residents retired or semi-retired, providing long-term stability to the community.
- Silent Generation: Representing 431 households, these residents are primarily nonworking, further cementing the community's mature demographic.
- Gen X & Younger Boomers: A rapidly expanding segment, these working residents—often in hybrid or remote roles—constitute 30% of respondents in the latest December 2023 survey, a sharp increase from less than 10% in 2021 and 5% in 2018.
- Parental & Multi-Generational Households: Though relatively small at 100 households, this segment highlights a growing trend toward more diverse family structures.
- Filial Households: Comprising 125 households, these arrangements indicate an increasing presence of adult children caring for older relatives.
- Millennial-Led Households: With only 8 instances, this group remains a small minority, reaffirming the community's predominantly older demographic.

Consistent with the 2018 survey, approximately 11% of households consist of single residents. This figure is lower than census estimates, suggesting that elderly widows and widowers may be underrepresented among survey respondents.

While Tellico Village remains anchored by an aging majority, the rise of younger, working residents and evolving household structures signals a gradual demographic shift. These demographic patterns set the stage for variations in tenure, as explored in the following section.

**Tenure Patterns**

Survey-based tenure estimates provide insight into homeowner stability but are subject to response biases, particularly underrepresentation of older residents. The 2018 and 2021 large-scale surveys show reasonable alignment in median tenure estimates for survey responders, with the 2018 binned estimate (9.72 years) closely matching the 2021 raw data estimate (8.35 years) when accounting for standard error differences. However, the 2021 binned estimate (14.43 years) is notably higher, likely inflated by the broad "20+ years" category, which introduces additional variability.

Further analysis reveals significant underrepresentation of the 80+ age group, leading to systematic underestimation of actual tenure in survey results. Kaplan-Meier survival analysis of Property Appraiser records confirms that homeownership duration increased from ~4 years in 2000 to 13 years in 2020, with a statistically significant decline to 11 years post-COVID ($p = 3.3e-22$, log-rank test).

Figure 1 presents Kaplan-Meier survival curves, illustrating the probability of a homeowner remaining in their residence over time across different neighborhoods. The median tenure time—the point at which half of the residents have sold their homes—is estimated from these survival curves. Notably, tenure distributions vary significantly across neighborhoods, with some showing steeper declines in homeownership retention than others. Additionally, not all neighborhoods have data extending beyond 35 years, as only the oldest neighborhoods (A, B, and C) have a long enough history to reach that threshold.)

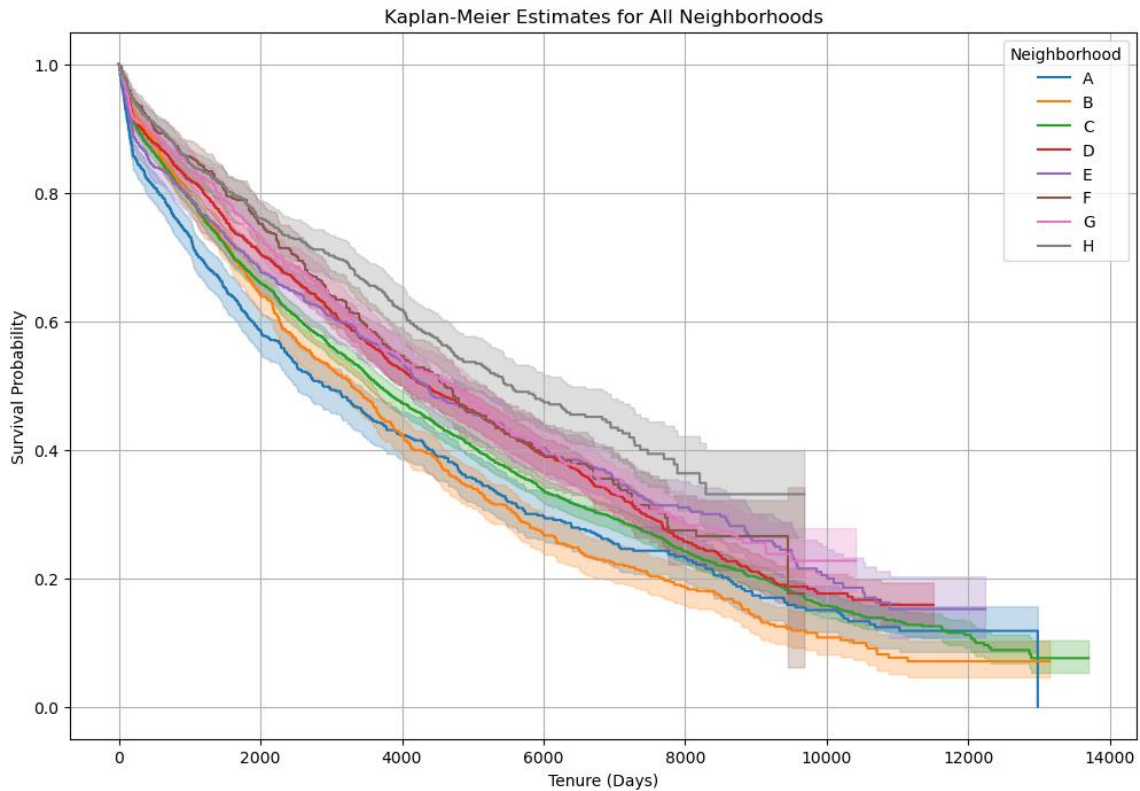

*Figure 1: **Kaplan-Meier Survival Curves for Homeowner Tenure Across Neighborhoods***
*The figure illustrates the probability of homeowners remaining in their residences over time for eight neighborhoods (A–H). Median tenure—the point at which half of the residents have sold their homes—is estimated from these survival curves. Only the oldest neighborhoods (A, B, and C) have data extending beyond 35 years, reflecting their longer historical records.*

Neighborhood-level analysis further reveals distinct tenure clusters. ANOVA results indicate statistically significant differences ($p < 0.001$), and post-hoc tests suggest neighborhoods fall into three tenure groups:

Short Tenure Neighborhoods: A, B, C
Medium Tenure Neighborhoods: D, E, F, G
Long Tenure Neighborhood: H

These clusters highlight how different parts of the community experience distinct tenure dynamics. Pre- and post-COVID tenure shifts are detailed in Table 1, which presents Kaplan-Meier median tenure estimates for periods before January 2000 and through January 2025. The table also includes participation rates (percentage of households that responded to the survey vs. total neighborhood households), average satisfaction scores, recommitment rates (percentage of respondents who affirmed they would still choose to live in the community), and the COVID Impact Index (CII).

*Table 1 **Neighborhood Tenure, Satisfaction, and COVID Impact Summary.** The first column lists eight neighborhoods (A–H), with survey response rates in parentheses (percentage of neighborhood population responding in 2021). The last row, ALL, represents overall community values. Second and subsequent columns present: Average satisfaction scores (5-point Likert scale: from 1 = very dissatisfied to 5 = very satisfied); Recommitment rates (percentage of residents who would still choose the community if given the choice again), Pre-COVID and post-COVID median tenures (according to Kaplan-Meier estimates), relative tenure change, p-values for tenure shifts, App, average market appraisal in millions and CII, COVID impact index, as previously defined.*

| Neighborhood (% response) | Average Satisfaction | Recommit-ment | Pre-COVID | Post-COVID | Relative Change | p-value | App | CII |
|---|---|---|---|---|---|---|---|---|
| A (48%) | 4.2 | 84% | 11.93 | 7.89 | 0.34 | 2.00E-06 | 0.40 | 0.3 |
| B (47%) | 4.3 | 88% | 10.56 | 8.99 | 0.15 | 1.90E-05 | 0.43 | 0.1 |
| C (41%) | 4.4 | 94% | 11.96 | 10.04 | 0.16 | 0.244 | 0.45 | 0.1 |
| D (48%) | 4.4 | 92% | 15.09 | 11.84 | 0.22 | 2.85E-07 | 0.53 | 0.2 |
| E (49%) | 4.3 | 90% | 15.7 | 11.84 | 0.25 | 1.11E-07 | 0.54 | 0.2 |
| F (48%) | 4.3 | 88% | 14.91 | 12.61 | 0.15 | 0.00557 | 0.71 | 0.1 |
| G (52%) | 4.4 | 91% | 14.64 | 12.9 | 0.12 | 0.00228 | 0.57 | 0.1 |
| H (71%) | 4.4 | 94% | 20.46 | 15.19 | 0.26 | 0.00298 | 0.73 | 0.1 |
| ALL (49%) | 4.3 | 91% | 13.03 | 10.63 | 0.18 | 3.3E-22 | 0.52 | 0.2 |

As shown in Table 1, neighborhoods A and H represent the extremes, with one of the oldest neighborhoods, A, exhibiting the lowest tenure and satisfaction, while the newest neighborhood H shows the highest values. However, tenure and satisfaction trends in the intermediate neighborhoods are not uniformly distributed.

To assess key predictors of tenure, we conducted a hedonic regression analysis, standardizing predictors for fair comparison. Variance Inflation Factors (VIF) were calculated to assess multicollinearity, and an Ordinary Least Squares (OLS) regression was estimated using robust standard errors (HC3) to account for heteroscedasticity. Variables were standardized using z-scores (mean = 0, standard deviation = 1) before running regressions.

The appraisal variable (average appraisal values computed from appraisals of all homes in the neighborhood) had the highest standardized coefficients, indicating its dominant influence on tenure. Consistently, predictive strength ($R^2$ scores) also showed that appraisal values ($R^2 = 0.808$) were the strongest predictor of tenure, more predictive than percentage of houses with private boat slips in the neighborhood (($R^2 = 0.414$), overall satisfaction (0.405), golf satisfaction (0.393) and recreational satisfaction (0.232) that contribute to a much lesser extent. Unlike appraisal values, satisfaction does not increase linearly with tenure but varies, suggesting other influencing factors.

Figure 2 Shows Median Tenures in the 8 neighborhoods (how long residents stay in a neighborhood) vs average Appraisal values. Red Regression Line shows the overall trend, indicating a positive correlation—higher tenure is associated with higher appraisal values.

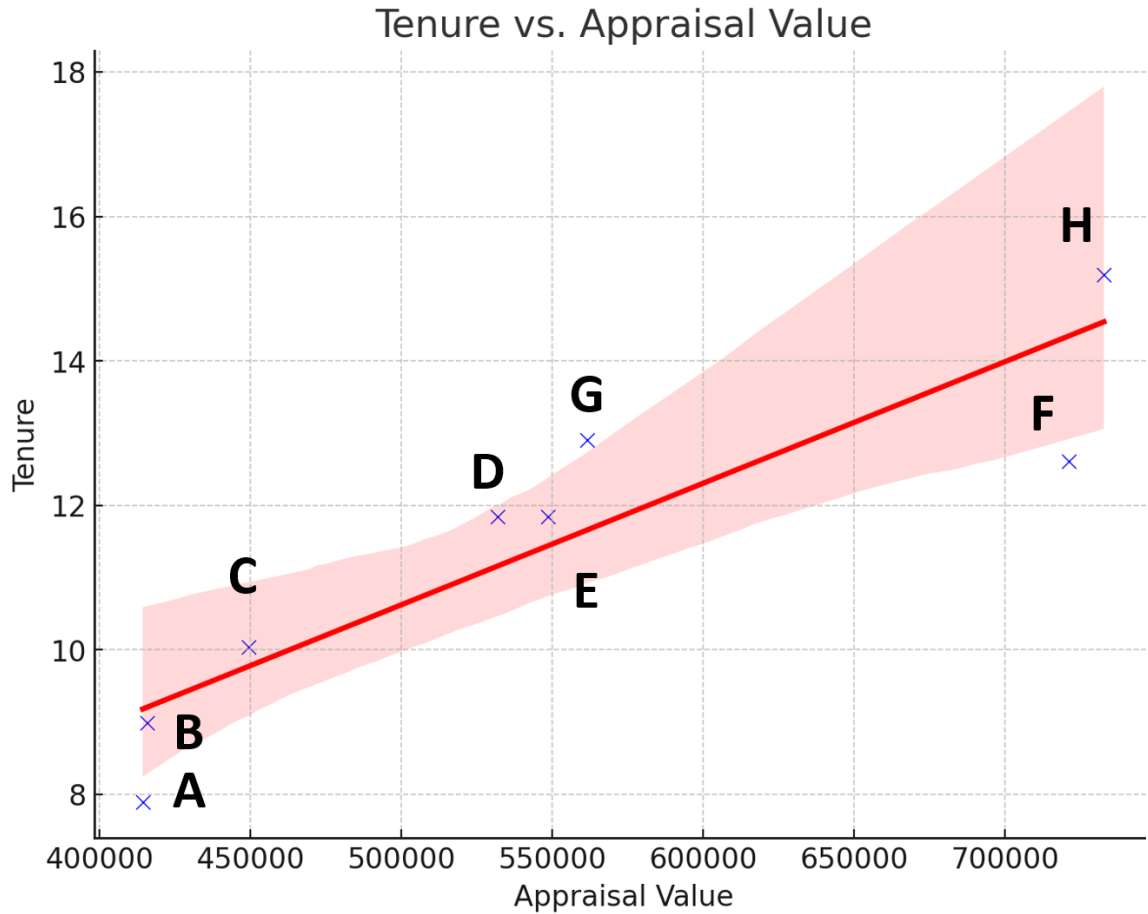

*Figure 2 **Relationship Between Median Tenure and Average Appraisal Values Across Neighborhoods A-H.** The red regression line highlighting the positive correlation—higher tenure is associated with higher appraisal values. The pink shaded area represents the confidence interval, which provides an estimate of the uncertainty around the regression line—wider areas indicate more variability in the data.*

Beyond neighborhood-based differences, deeper analysis of home sales trends and turnover dynamics reveals additional tenure complexities.

**Turnover Dynamics**

Resident turnover follows distinct patterns driven by both personal life transitions and external market factors. Histograms of home sales over time reveal multiple tenure segments, including:

Ultra-short-term property owners (mostly builders, ~1 year)
Short-term residents (3–9 years)

Long-term residents (10+ years)

A decomposition of tenure durations using Weibull distributions highlights that while a stable long-term segment remains, its relative size is shrinking. A noisy but observable pattern in the histograms suggests multiple tenure peaks, shifting slightly depending on sales year.

Hazard rate analysis—which examines the likelihood of departure on any given day, conditional on continued residency—identifies multiple critical departure periods, including pronounced spikes at 2.7 years (Day 969, 1.5% hazard rate) and 25.9 years (Day 9446, 1.2% hazard rate). However, due to the inherent volatility in daily hazard rates, a yearly analysis framework was introduced to stabilize long-term movement patterns and mitigate short-term fluctuations that could obscure underlying trends.

As shown in Figure 2, this smoothed approach reveals key departure peaks at 5, 7, 11, 16, 22, and 26 years. Notably, data beyond 26 years becomes increasingly noisy due to the community's relatively short history—founded on December 15, 1985, and experiencing only moderate growth in its early years.

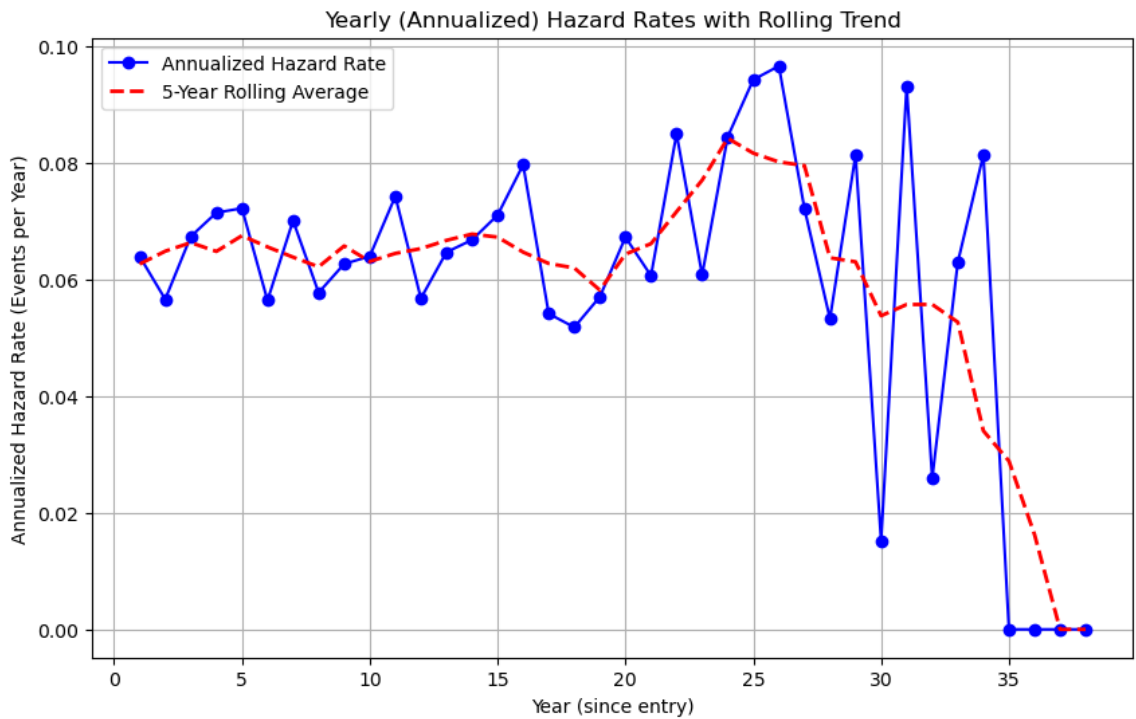

*Figure 3* ***Annualized Hazard Rates Over Time with Rolling Trend***
*This figure illustrates yearly (annualized) hazard rates, representing the probability of an event (household's departure from the community) occurring per year since entry. The blue line with markers shows the annualized hazard rate for each year, calculated by aggregating daily hazard rates into yearly bins. The red dashed line represents a rolling average trend using a 5-year window, smoothing fluctuations to highlight long-term patterns.*

Annual sales histograms further validate these patterns, highlighting major departure peaks at 5 (3-5), 7, 11, 16 and 26 years. Additionally, entry-year analysis indicates a large influx of residents during the 2005–2006 housing boom, followed by larger than average exits 10–15 years later, reinforcing the role of economic cycles in shaping tenure dynamics.

While most departures represent residents leaving the community entirely, a small fraction relocate within Tellico Village. Only about 2% of cases involve current residents building new homes, meaning this likely has minimal impact on overall tenure statistics. Among internal relocators, some downsize, others move to a different neighborhood, and some merge households, such as bringing in aging parents.
In contrast to The Villages—the largest active adult community, where residents relocate an average of 2.5 times across its many neighborhoods, the smaller, more stable nature of Tellico Village naturally results in significantly fewer internal moves, which is consistent with its overall design and scale.

### Neighborhood-Level Variations and Engagement

Our analysis reveals that each neighborhood displays distinct patterns of tenure, resident satisfaction, and community engagement, influenced by demographic composition, infrastructure, and local economic conditions. Below is a detailed examination of each neighborhood's strengths and weaknesses, along with possible drivers for the observed trends.

#### Neighborhood A

**Strengths:** One of the oldest neighborhoods in the village close to retail. Rapid post-2020 construction growth suggests ongoing development and potential for revitalization.
**Weaknesses:** Residents experienced the greatest pandemic impact (see Table 1), leading to the shortest post-COVID tenure, lowest satisfaction, and minimal recommitment.
A high proportion of single households and low willingness to carpool may signal weaker social ties and reduced community cohesion.
Economic indicators lag, as reflected in the lowest property appraisals and appreciation rates, and there is diminished interest in healthy restaurant options.
**Possible Drivers:** A combination of rapid development and a transient population, particularly among single households with lower social cohesion, may contribute to lower long-term commitment and community satisfaction.

#### Neighborhood B

**Strengths:** A clear preference for townhome living suggests a strong market niche, particularly for residents envisioning aging in place.
**Weaknesses:** Shorter pre- and post-COVID tenures indicate a more transient population.

Lower appreciation for natural amenities (lake and mountain views), recreational activities (golf, boating, kayaking), and minimal civic participation may contribute to weaker community identity.
**Possible Drivers:** The focus on townhome living, without a strong attachment to local natural features or recreational activities, could explain lower civic engagement and satisfaction.

### Neighborhood C

**Strengths:** One of the highest average satisfaction levels underscores a robust sense of well-being among residents. Engagement is high, evidenced by detailed and lengthy general comments and a clear appetite for urban conveniences, such as commercial businesses and restaurants.
**Weaknesses:** Limited interest in communal meeting spaces may indicate a preference for informal interactions or a desire for more modern, flexible gathering options.
**Possible Drivers:** The community's prioritization of urban amenities and commercial vitality, cartability (golf-cart friendliness) and amenities inside the neighborhood likely fuels higher satisfaction, while the focus on driving range expansion points to evolving recreational interests.

### Neighborhood D

**Strengths:** Strong pre-COVID tenure and high satisfaction suggest stability and a well-rooted community. A pronounced valuation of natural beauty as a contributor to property values reinforces a sense of place and quality of life. The neighborhood leads in participation in large events and shows significant interest in carpooling and ride share services, highlighting strong community engagement and connectivity.
**Weaknesses**: Frequent complaints about parking and event capacity hint at infrastructure challenges that may hinder community activities.
Lower trust in governance—particularly regarding POA-operated businesses—could limit the effectiveness of future community initiatives.
**Possible Drivers**: Established tenure, a focus on natural amenities, and active community participation (e.g., in large events and shared mobility) contribute to overall satisfaction. However, persistent infrastructure issues like parking and event capacity might detract from the community's appeal over time if not addressed.

### Neighborhood E

**Strengths:** A vibrant boating culture is evident, with the highest numbers of boat and kayak owners and a significant percentage of private boat slips. A strong focus on financial priorities among households may reflect an engaged and economically savvy community.
**Weaknesses:** Midrange satisfaction levels and a larger proportion of neutral responses suggest ambivalence toward community offerings. Limited interest in downsizing or alternative housing options, combined with a highly built-out landscape, might constrain long-term adaptability.
**Possible Drivers:** Economic considerations and a mature development pattern seem to shape residents' priorities, yet a lack of housing flexibility and moderated satisfaction levels indicate potential challenges for future community evolution.

### Neighborhood F

**Strengths:** Stable Community Base: The neighborhood benefits from a stable demographic profile, with a significant proportion of older long-term residents, mostly couples. This can foster strong community bonds and consistent neighborhood identity. Support for Predictable Fee Increases: Residents show the highest support for predictable annual fee increases, which can ensure steady funding for essential maintenance and community services.
**Weaknesses:** High levels of parking complaints indicate that the current infrastructure may not adequately meet residents' needs, particularly for the aging population. This issue could affect overall resident satisfaction and daily convenience. Limited Engagement in Recreational Offerings like fitness classes might reflect a disconnect between the available amenities and the preferences of the primarily older demographic. Low Interest in On-Demand Services like Uber-like services suggest a potential misalignment between service offerings and resident needs or preferences.
**Possible Drivers:** The predominance of older residents coupled with infrastructural shortcomings (especially related to parking) could lead to reduced satisfaction and lower engagement with new or existing amenities. There is an observable trend where residents who participate in recreational activities tend to report higher satisfaction. However, Neighborhood F has fewer working-age individuals and single older residents who might not find the current community offerings appealing. This imbalance might contribute to a sense of disconnect for these groups, potentially affecting overall community cohesion.

### Neighborhood G

**Strengths:** Despite traffic congestion issues, this neighborhood boasts the highest number of "very satisfied" residents and a unique status as entirely owner-occupied. A niche community of golf connoisseurs suggests a concentrated, high-end recreational market. High interest in meeting spaces aligns with the neighborhood's status as one of the oldest, with an average resident age only slightly lower than Neighborhood F.
**Weaknesses:** Persistent traffic complaints may indicate underlying transportation and infrastructure challenges. Highest interest in meeting spaces. The low percentage of renters might limit diversity and potentially reduce the influx of new perspectives.
**Possible Drivers:** Strong owner-occupancy and specialized recreational interests contribute to high satisfaction, despite infrastructure concerns like traffic congestion.

*Neighborhood H*
**Strengths:** Exceptional social cohesion is reflected in active participation in events, groups, and online networks, alongside the highest median tenure and largest family sizes. Strong community engagement is evident in broad support for initiatives such as healthy food options and housing diversification, aligning with high recommitment rates (residents who would choose the community again) and satisfaction levels. Robust economic performance, characterized by high property appraisals and strong appreciation rates, reinforces long-term desirability.
**Weaknesses:** While generally positive, high support for increased property owner dues may indicate underlying financial pressures, which could become a challenge if not balanced with service enhancements. Lowest support for increasing user fees and private funding contributions, despite the community's history of substantial private investment, suggests a shift in funding preferences—residents favor shared cost increases (higher dues) over additional personal contributions.
**Key Drivers:** Deep-rooted social networks and a stable, family-oriented demographic drive long-term commitment and satisfaction. Economic success further strengthens these trends, though balancing funding strategies (e.g., shared dues vs. user fees) remains a key policy consideration.

Across neighborhoods, several key themes emerge:

Demographic Composition:

The blend of age, household type, and homeownership versus renting consistently shapes tenure and satisfaction. For instance, family-oriented, mostly owner-occupied communities (like Neighborhoods G and H) generally report higher satisfaction and commitment. Notably, even though Neighborhood F, with its older population, shows strong tenure, overall satisfaction is not as high—indicating that while older communities may demonstrate long-term commitment, this does not automatically translate into higher satisfaction.

Infrastructure and Amenities:

The quality and accessibility of local amenities—from natural beauty and recreational options to parking and commercial services—play a critical role in resident contentment. Neighborhoods with robust recreational facilities and well-developed infrastructure (such as H, D, and C) tend to be more satisfied. In contrast, areas undergoing rapid construction (like A) or facing infrastructural challenges, such as parking issues (as seen in F), experience more resident dissatisfaction.

Economic Forces:

Property appraisals, development patterns, and resident financial priorities are closely linked to satisfaction trends.  Property appreciation is another key factor in resident satisfaction (highest in H, lowest in A, see table 1). Communities featuring on-site amenities such as restaurants and golf courses tend to see higher property values and more satisfied residents. Notably, the concept of "cartability" (C, D and H) plays an important role here. When residents can easily visit local restaurants and amenities by walking or using a golf cart—without the need to drive on highways to neighboring areas—it significantly enhances their overall experience. This ease of access not only boosts satisfaction but can also contribute to higher property appreciation. Additionally, such integrated amenities can facilitate private funding initiatives, further reinforcing community development and economic stability.

Community Engagement:

Social cohesion and civic participation significantly influence residents' perceptions of their neighborhoods. Highly engaged communities, such as Neighborhood H, show stronger rates of recommitment. Conversely, neighborhoods with lower civic participation (like B) are more likely to experience transient residency and reduced overall satisfaction.

Tailored policies that address affordability, transportation improvements, and amenity development—while taking into account the unique profile of each neighborhood—can enhance resident satisfaction and promote sustainable community engagement.
By considering these interrelated factors, stakeholders can better understand the dynamics of neighborhood satisfaction and devise strategies to foster more resilient, connected, and content communities.

**Resident Satisfaction and Key Predictors**

Our analysis shows that resident satisfaction in active adult communities is closely linked to financial attitudes, community engagement, and neighborhood characteristics. Key findings include:

There is an almost linear relationship between willingness to pay higher fees and overall satisfaction—each incremental increase in willingness corresponds to a statistically significant rise in satisfaction. The Kruskal-Wallis test ($p = 8.9\text{E}{-}25$) confirms a strong overall effect, while all pairwise Mann-Whitney tests ($p < 0.01$) support significant differences, except between the two groups with the highest willingness to pay. The $R^2$ is 94% (see Appendix and Figure 4).

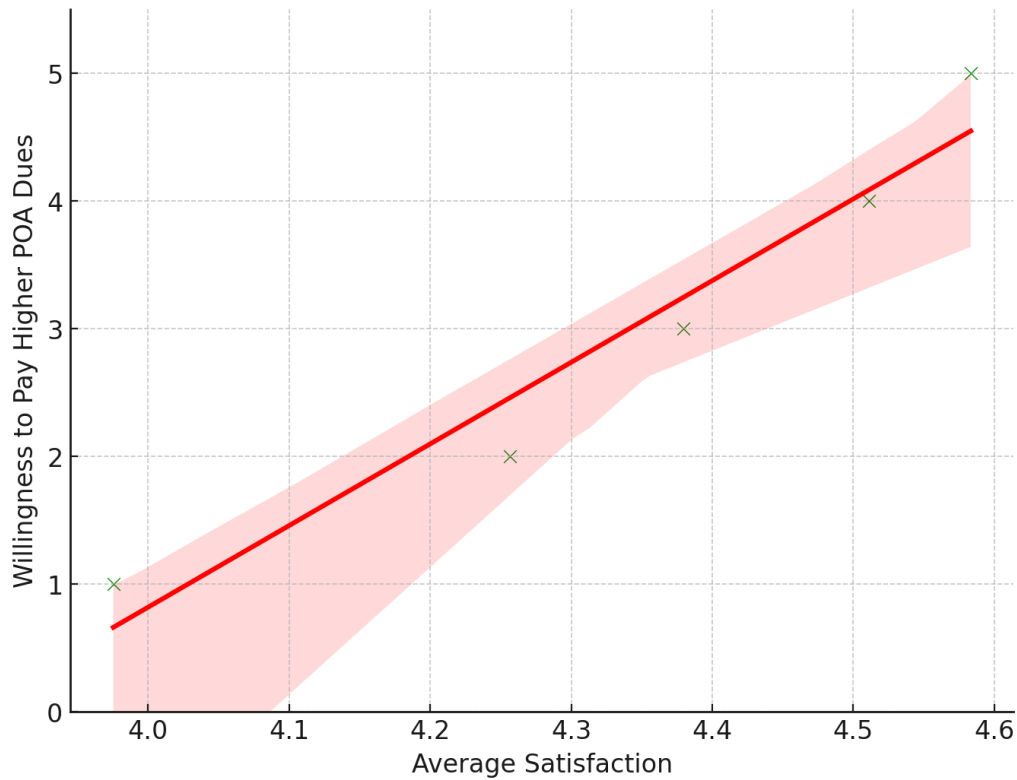

*Figure 4 **Relationship between average community satisfaction** and different levels of willingness to pay higher POA dues across all residents. Each point represents a unique willingness level (1, 2, 3, 4, 5) and its corresponding average satisfaction score across the entire community. The red regression line highlights the near-perfect linear trend, with 94% of the variation in satisfaction is explained by willingness to pay higher dues. The pink shaded area represents the confidence interval, showing minimal uncertainty due to the aggregation of responses, which reduces noise in the data.*

Neighborhood-level averages indicate that residents who support structured fee increases tend to be more satisfied, whereas additional user fees correlate with lower satisfaction.

Generational differences are significant: Boomers and Silents have consistently shown higher willingness to pay compared to GenX, Millennials, and GenZ. Although Silent generation's willingness slightly declined in 2021 (as expected in advanced ages), it remains statistically higher than that of younger cohorts (Fig.5).

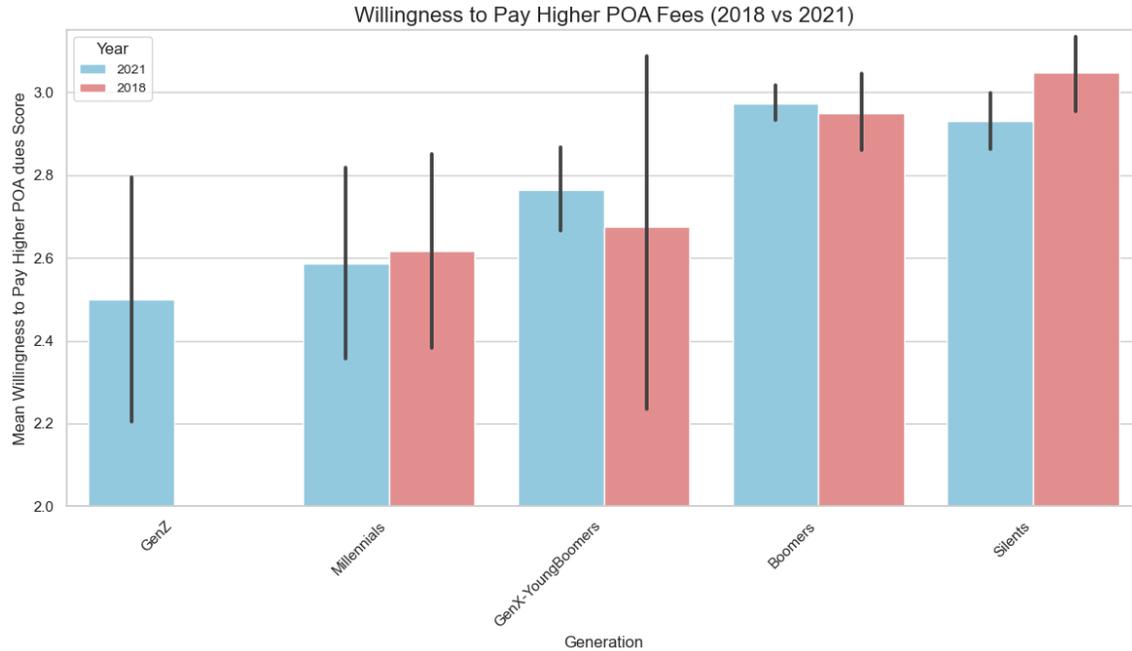

*Figure 5 **Willingness to Pay Higher Fees by Generation (2018 vs. 2021)***
*Generational willingness to pay higher fees is presented for the 2018 and 2021 survey results, with error bars representing the standard deviation*

Additionally, recommitment rates among millennial-supported households are lower than those with Generation Z residents. Statistical analysis supports these findings, with a Kruskal-Wallis test yielding a p-value of 0.0075 for recommitment likelihood and 0.0056 for satisfaction levels across different generational groups.

The 2018 survey data reveal that residents with less than five years of tenure exhibited significantly lower willingness to pay (WTP) than those with 11–20 years (pre-financial crisis cohort). However, no statistically significant difference was observed between other tenure groups. By 2021, WTP for residents with less than five years of tenure remained significantly lower than that of the 11–20-year group, but now also statistically significantly lower than residents of 5–10 years. WTP for 20+ years still did not significantly differ from those with less than five years.

In 2021, the newest residents (0–2 years) had replaced the longest-tenured (20+ years) as the least willing to pay, with an overall decline in WTP likely influenced by broader economic conditions and shifting demographics. By January 2025, more than half of the community had experienced turnover in the past six years, highlighting the importance of engaging new residents in shared community goals and illustrating the benefits of continued investment in upkeep.

Resident satisfaction follows a U-shaped trajectory relative to tenure, with the lowest satisfaction seen among residents with 6–9 years in the community (Figure 6)

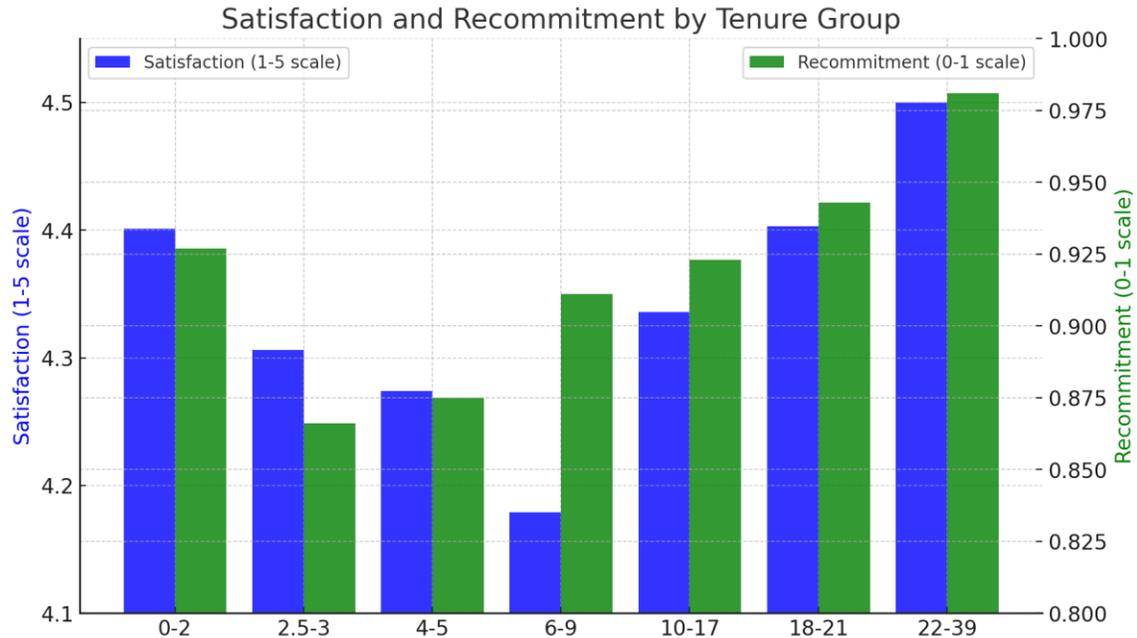

*Figure 6 **Satisfaction and Recommitment by Tenure Group**. Resident satisfaction follows a U-shaped trajectory relative to tenure, with the lowest satisfaction observed among residents with 5–9 (2.5–12) years in the community. The left y-axis (blue bars) represents satisfaction scores on a 5-point Likert scale (1–5), while the right y-axis (green bars) represents recommitment likelihood (binary responses, scaled between 0 and 1 for comparison).*

Decision tree binning techniques confirm that residents with tenures of 2.5–12 years exhibit statistically lower satisfaction than both newer (0–2 years) and longer-term (13+ years) residents. At the sub-neighborhood level, tenure-satisfaction relationships vary: in the oldest neighborhood (F), the satisfaction dip shifts to the 10–15-year range but subsequently recovers.

Based on US data on mortality and disability [9-10], starting at age 65 (average age of residents arriving to active lifestyle community), the median time until the composite event (disability or death) is roughly 13 years. Our data [11] show that Tellico Village is healthier than average. Hence, departure spikes at 16, 22 and 26 years (Figure 3) correlate with aging residents transitioning to assisted living or downsizing, with some exits driven by mortality.

Families with millennials report significantly lower satisfaction compared to those led by older generations. In terms of recommitment, their scores are even lower than households with Generation Z members (Figure 7). We note that there are no Gen Z-led households in the village; most Generation Z residents live with their parents, primarily from Generation X. The Kruskal-Wallis test indicates statistically significant differences in satisfaction ($p = 0.0056$) and recommitment ($p = 0.0075$) across generations. Satisfaction is a multifaceted function of amenities, maintenance, safety, social connections, cost, and location.

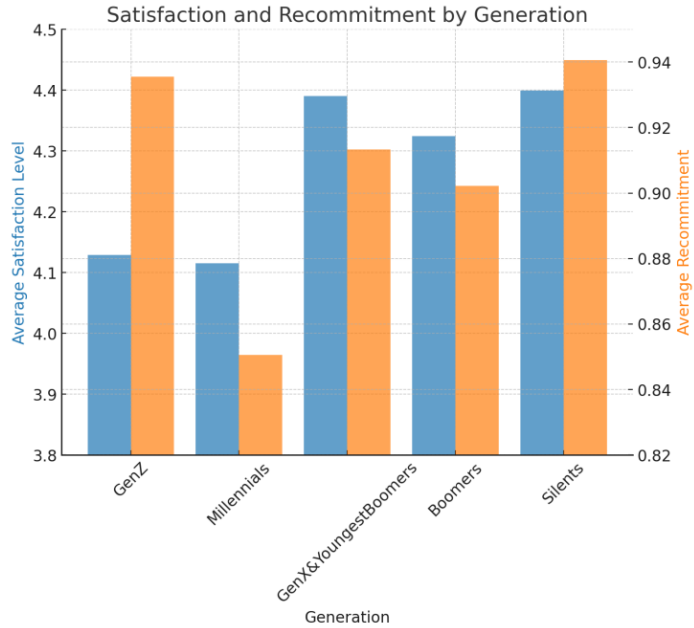

*Figure 7 Satisfaction and Recommitment by Generation. Satisfaction and recommitment levels are shown by generation, with the left y-axis representing satisfaction (scaled from 3.8 to 4.5) and the right y-axis representing recommitment (scaled from 0.82 to 0.95). Generational differences in these measures highlight trends in community engagement and long-term commitment.*

While neighborhood-level correlations (e.g., between satisfaction and support for fee policies or property appraisals) are robust, individual-level predictions remain challenging due to lower correlations and high variability.

Key individual predictors include aesthetics satisfaction followed by active recreational participation and willingness to pay higher monthly dues, though their impacts are less pronounced than broader trends.

Overall, resident satisfaction in these communities is shaped by a complex interplay of financial preferences, tenure, and community characteristics. These insights can guide community management strategies to balance amenity investments with the cost sensitivities of newer residents, ultimately enhancing long-term retention and satisfaction.

## Conclusion

While Baby Boomers and the Silent Generation remain the majority in active adult lifestyle communities, the increasing presence of Gen X and remotely-working residents signals a subtle yet meaningful demographic evolution. These shifts coincide with post-COVID tenure pattern changes and accelerated turnover.

Our findings confirm a post-COVID decline in homeownership duration, with median tenure falling from 13 to 11 years, aligning with national trends. Turnover follows structured patterns, with departure risk peaking at 3, 5, 7, 11, 16, 22, and 26 years—coinciding with personal life transitions, market cycles, and community adaptation dynamics. Notably, histograms of exits in recent years show increasing fractions of 5-, 7-, and 11-year tenure exits, highlighting accelerated mobility post-COVID.

This study integrates survival analysis, hazard modeling, and machine learning to provide a comprehensive view of tenure dynamics in active adult communities. Our findings highlight that resident turnover is influenced by both structural factors (e.g., community infrastructure, governance) and external conditions (e.g., economic shifts, public health events). Property appraisal values significantly impact tenure, with higher-valued neighborhoods exhibiting longer retention. However, satisfaction and commitment are also shaped by infrastructure adequacy, recreational opportunities, and social engagement.

Neighborhood-level variations in tenure, satisfaction, and recommitment rates underscore distinct community dynamics, informing broader trends in housing stability and quality of life. Our findings contribute to the emerging study of Neighborhood Dynamics in Active Lifestyle Communities, demonstrating that homeownership duration is shaped by both macroeconomic conditions (e.g., interest rates, recessions) and micro-level influences (e.g., life events, satisfaction, amenities). Resident satisfaction follows a U-shaped trajectory, where new and long-term residents report higher satisfaction, while mid-tenure residents experience declines, mirroring adaptation patterns observed in other domains.

Financial preferences, tenure dynamics, and neighborhood characteristics collectively shape satisfaction. Higher willingness to pay increased POA dues to cover emerging maintenance needs correlates with greater satisfaction, with Boomers and Silents more accepting of higher costs than younger cohorts. Additionally, social cohesion, amenity availability, and infrastructure quality play critical roles in shaping long-term retention and well-being. The interaction between social and physical neighborhood characteristics further impacts health outcomes and overall community stability.

Resident turnover exhibits structured patterns, with peaks aligning with life transitions, economic fluctuations, and community design. These findings emphasize the need for policies that align infrastructure investments with affordability, ensuring sustainable satisfaction and retention in active adult communities.

This research shifts the focus from undeveloped land transactions [7] to the lived experiences of residents in completed homes. By integrating large-scale community surveys with property transaction records, we uncover patterns of satisfaction, mobility, and attachment that have been largely unexplored. While neighborhood-level trends are robust, individual-level predictions remain complex due to high variability, underscoring the need for more granular, longitudinal data.

Ultimately, our findings provide a framework for optimizing community management strategies—balancing amenity investments with fee structures to accommodate cost-sensitive new residents while promoting long-term retention. Understanding these dynamics is essential for sustaining satisfaction and ensuring stability in active adult communities.

### Acknowledgements

We appreciate the coding support provided by ChatGPT 4o, o1 and o3.

**Funding**

This research received no external funding.

**Data Availability Statement**

This study was conducted as an open-source collaborative project through the [Open Science Framework](), registered under DOI 10.17605/OSF.IO/TGV6Q. Data is available at OSF: [https://osf.io/w7682/s](). Associated code is available at [GitHub](): https://github.com/MPC-Dynamic/Growth

**Conflicts of Interest**

None declared.

## Abbreviations and Terms used in this paper

AALC: Active Adult Lifestyle Community – A residential community offering amenities and services tailored to an active retirement lifestyle.
ANOVA: Analysis of Variance – A statistical method used to compare means across multiple groups when the data is normally distributed. If normality assumptions are violated, the Kruskal-Wallis test is used as a non-parametric alternative.
Cartability: golf-cart driveability, similar to "walkability
CDC: Centers for Disease Control and Prevention
CII: COVID Impact Index
LASSO: Least Absolute Shrinkage and Selection Operator – A regression technique that performs both variable selection and regularization to enhance model prediction accuracy by reducing overfitting.
MPC: Master-Planned Community – A large-scale residential development with carefully designed infrastructure, amenities, and land use planning to create a cohesive living environment.
MT: median tenure, the time point at which 50% of households had moved out
ACS: American Community Survey – A U.S. Census Bureau survey that provides demographic, economic, and housing data at the national, state, and community levels.
NCHS: National Center for Health Statistics – A division of the CDC responsible for collecting and analyzing health-related data to guide public health policies and programs.
POA: Property Owners Association – A governing body that oversees common property, amenities, and community regulations within a residential development.
Recommitment: The percentage of survey respondents who indicated that they would choose to live in the community again if given the choice.
SVM: Support Vector Machines – A supervised machine learning particularly effective in high-dimensional spaces and with imbalanced data.

TPAD: Tennessee Property Assessment Data – A dataset containing property valuation, ownership, and assessment details for real estate in Tennessee, often used for market analysis and research.
Tukey's HSF: Tukey's honest significance test, a single-step multiple pairwise comparison procedure
WTP – willingness to pay higher POA dues
XGBoost: Extreme Gradient Boosting – A machine learning algorithm that optimizes decision trees for high predictive accuracy

# Appendix

### Differences between satisfaction levels across willingness-to-pay groups

Group1: Score 1 – "Ultra Frugal": "strongly support" for keeping monthly assessments at a minimum and "do not support" annual fee increases. It reflects an extremely cost-sensitive stance where only the barest essential costs are acceptable, with no appetite for extra spending.
Group 2: Score 2 – "Cautiously Conservative"
Whether from favoring minimal fees (but only "supporting" a review) or from a moderate stance on monthly fees (with "no support" for increases), these respondents lean strongly toward low, fixed costs. They're very cautious about any extra charges and prefer to keep community fees as low as possible—even if that means only minor flexibility in annual reviews.
Group 3: Score 3 – "Balanced Budgeter"
This middle score can arise in a few ways—for example, by strongly favoring minimal monthly assessments but also strongly supporting annual reviews, or by a mix of moderate positions. It indicates a balanced view: these residents value cost control but also recognize that occasional fee adjustments may be necessary to maintain the community's long-term needs.
Group 4: Score 4 – "Maintenance-Minded"
With a score of 4, residents show a greater willingness to invest in the community. They are less fixed on keeping monthly fees at a minimum (or even oppose the "minimal fee" stance) and at least "support" annual reviews and adjustments. This reflects a proactive attitude toward ensuring that operations, maintenance, and amenity improvements are adequately funded.
Group 5: Score 5 – "Community Investor"
The highest score is reserved for those who oppose minimal monthly fees and "strongly support" annual fee reviews and adjustments. These residents are the most open to paying higher or more dynamic assessments, viewing them as necessary investments for robust operations, proactive maintenance, and future improvements.
The Shapiro-Wilk test indicates that the satisfaction levels do not follow a normal distribution (p-value < 0.01, highly significant deviation from normality). Because of this, Kruskal-Wallis H and Mann-Whitney U tests are more appropriate than t-tests for pairwise comparisons. The Kruskal-Wallis test confirms significant differences in satisfaction levels across willingness-to-pay groups. Test statistic is 118.97 and p-value is 8.9E-25. Mann-Whitney scores for pairwise comparisons are listed below.

| Group 1 | Group 2 | T-test p-value | Mann-Whitney p-value |
| --- | --- | --- | --- |
| 1 | 2 | 8.27E-06 | 2.80E-05 |
| 1 | 3 | 1.60E-11 | 1.05E-12 |
| 1 | 4 | 7.74E-17 | 1.74E-17 |
| 1 | 5 | 3.91E-19 | 1.00E-18 |
| 2 | 3 | 0.002085 | 0.000967 |
| 2 | 4 | 2.12E-08 | 2.21E-08 |
| 2 | 5 | 9.35E-11 | 2.15E-10 |
| 3 | 4 | 0.000762 | 0.001253 |
| 3 | 5 | 5.00E-06 | 1.27E-05 |
| 4 | 5 | 0.142074 | 0.136844 |

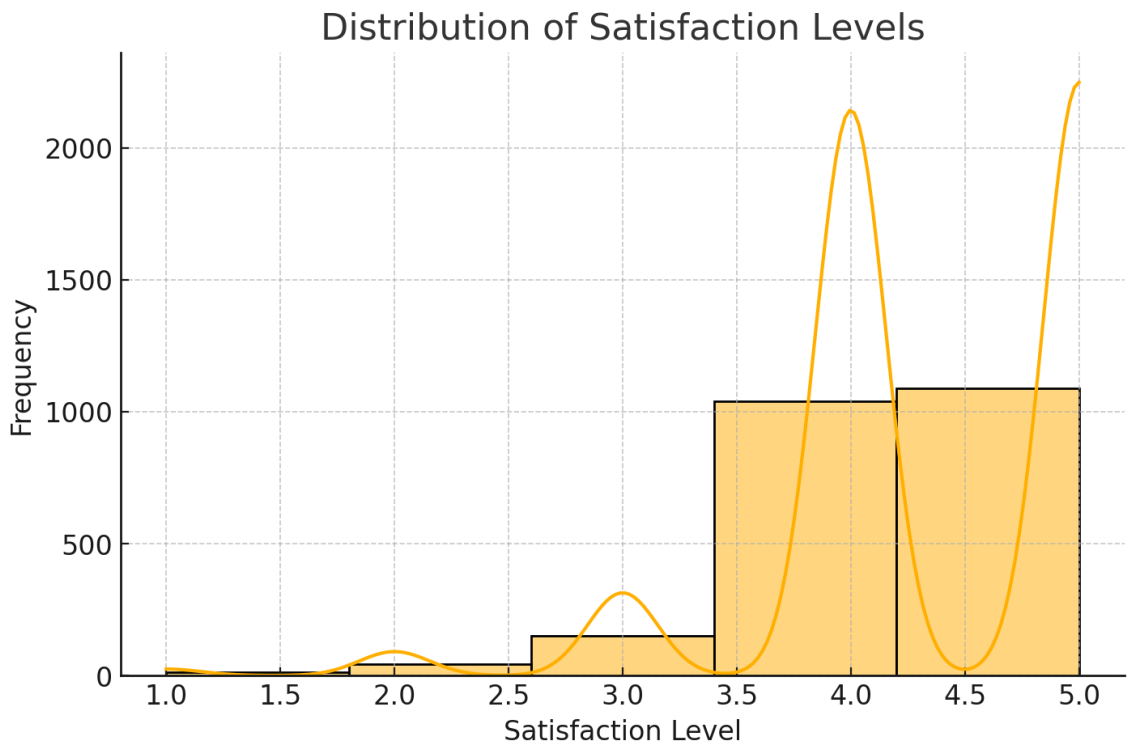

*Figure 8 The histogram with a KDE (Kernel Density Estimate) curve represents the distribution of overall community satisfaction levels. The data is on a 5-point Likert scale, meaning there are only five possible discrete values (1 to 5). The distribution is right-skewed, with more responses clustering around higher satisfaction levels*

At the individual household level, the model's $R^2$ of 0.0517 means that just 5.17% of the variation in satisfaction is explained by a household's willingness to pay higher POA dues. Although this relationship is statistically significant, it suggests that many other factors play important roles in determining satisfaction.

At the individual level, many other factors influence satisfaction, which dilutes the explanatory power of willingness to pay alone. However, when data is aggregated, these extraneous individual differences tend to cancel out, highlighting the primary trend. When we look at neighborhood averages, the model explains about 43.8% of the variance in satisfaction based on the average willingness to pay higher POA dues in each neighborhood.

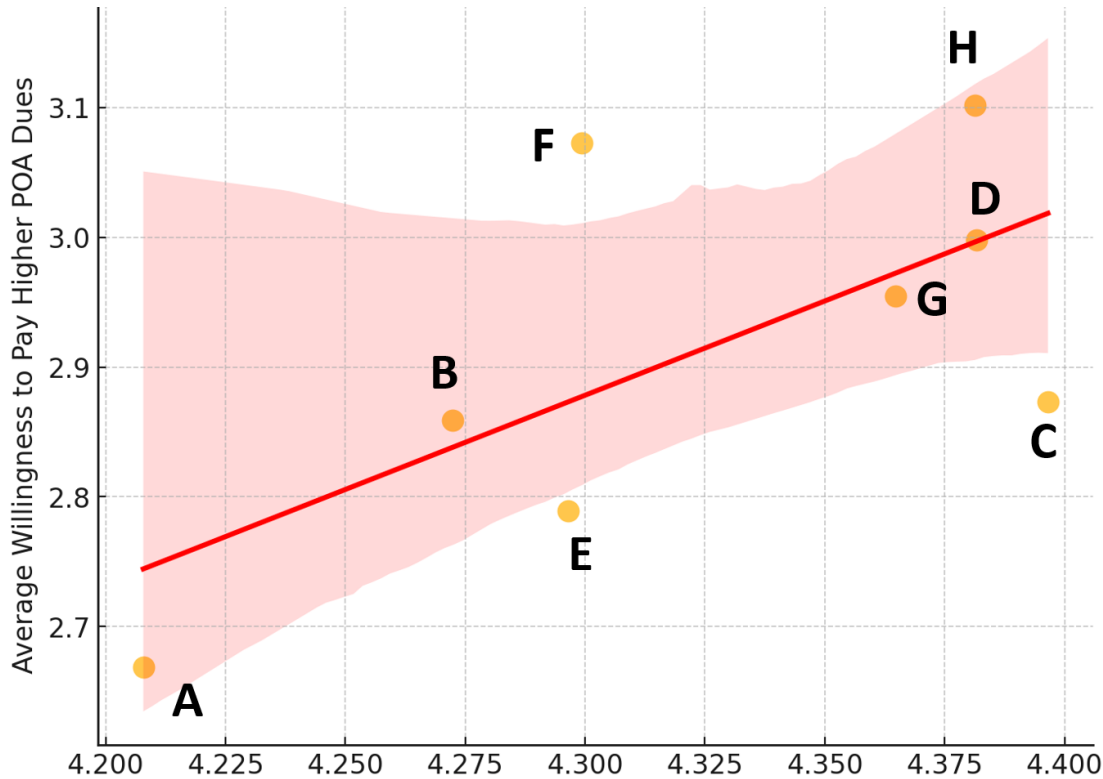

*Figure 9 Relationship Between Average Satisfaction and Willingness to Pay Higher POA Dues Across Neighborhoods A-H.*

This scatter plot with a regression line illustrates the relationship between average satisfaction (X axis) and average willingness to pay higher POA dues (Y axis) for each neighborhood. The correlation coefficient is 0.66, indicating a moderate positive relationship—residents with higher satisfaction levels are generally more willing to contribute higher POA dues.

When households are grouped by each specific level of willingness to pay higher POA dues, the model's $R^2$ increases to 94% (Figure 4).